\documentclass[prb,twocolumn]{revtex4}

\usepackage{graphicx}

\begin{document}

\title{
Quantum Disordered Ground States in Frustrated Antiferromagnets
with Multiple Ring Exchange Interactions
}

\author{Satoshi Fujimoto} 
\affiliation{
Department of Physics,
Kyoto University, Kyoto 606-8502, Japan
}

\date{\today}

\begin{abstract}
We present a certain class of two-dimensional frustrated quantum 
Heisenberg spin systems with multiple ring exchange
interactions which are rigorously demonstrated to 
have quantum disordered ground states without
magnetic long-range order. 
The systems considered in this paper are s=1/2 antiferromagnets on a honeycomb and 
square lattices, and an s=1 antiferromagnet on a triangular lattice.
We find that for a particular set of parameter values, the ground state is
a short-range resonating valence bond
state or a valence bond crystal state.
It is shown that these systems are closely related to the quantum dimer 
model introduced by Rokhsar and Kivelson as an effective low-energy theory
for valence bond states.
\end{abstract}


\maketitle

\section{Introduction}

Quantum frustrated Heisenberg antiferromagnets 
have attracted a great deal of interest
in connection with the search for exotic phases, such as spin liquids.
Anderson proposed a resonating valence
bond (RVB) state as a prototype of spin liquids
three decades ago~\cite{and}.
Since then, a number of works have studied the conjecture
that both geometrical frustration and 
strong quantum fluctuation
may destroy magnetic long-range order and stabilize a quantum disordered
liquid-like ground state without symmetry breaking~\cite{py,iso,moe,yama,koga,tsune,fuji,cha,sac,vdh,rit,lech,mila,zeng,mis2,lim}.
For example, it is believed quite likely that
quantum spin systems on Kagome lattices 
and pyrochlore lattices can exhibit
quantum spin liquid states, though
the elucidation of their ground states remains an important unsolved problem.
In related works, the quantum dimer model (QDM) introduced by
Rokhsar and Kivelson has been extensively
studied as an effective low-energy theory of
quantum disordered states with 
short-range antiferromagnetic correlation 
and a spin excitation gap~\cite{rk,suth,moe1,moe2,mis}.
The important feature of the QDM is that its Hilbert space is 
spanned by only dimer covering states
composed of singlet pairs of nearest-neighbor spins,
and this is the key to its success in 
describing quantum disordered states~\cite{rk}.
It was shown in the pioneer paper by Rokhsar and Kivelson that 
a short-range RVB state is realized in the QDM as its exact ground state.
It was also argued by several authors 
that the QDM may be an effective low-energy theory 
of quantum antiferromagnets on the kagome lattice.

However,
it is not unclear how to derive the QDM directly from
the original quantum spin Hamiltonian by truncating the Hilbert space.
Indeed, it appears difficult to find a general answer to this question.
Therefore, in this paper, we do not seek such a general result.
Instead, to obtain better insight regarding the mechanism stabilizing quantum
spin liquids, we attempt to make progress toward determining 
what kinds of quantum spin systems possess 
a low-energy sector described by the QDM.

The main result of this paper is the rigorous proof
that a certain class 
of quantum antiferromagnetic spin systems in two dimensions
with multiple ring exchange interactions
is equivalent to the QDM. 
These systems exhibit a short-range RVB state or a valence bond crystal (VBC) 
state, depending on the values of their parameters.
Our results clarify the important role played by 
the ring exchange interactions in the realization of 
quantum spin liquid states~\cite{bfg}.

This paper is organized as follows.
In Sec.II, we present results of an $s=1/2$ 
Heisenberg spin system with multiple ring exchange interactions
defined on a honeycomb lattice.
It is shown that for a particular parameter the rigorous ground state is
 the short range
RVB.
In this RVB state, spin-spin correlations decay exponentially, and therefore
there is no magnetic long range order.
On the other hand, correlations between spin-singlet dimers
exhibit power law long-distance behavior, implying
the existence of low-lying spin-singlet gapless excitations.
In Sec.III and Sec.IV, we consider an $s=1/2$ model on
a square lattice and an $s=1$ model on a triangular lattice, respecticely.
Discussion and summary are given in Sec.V.

\section{$s=1/2$ Heisenberg antiferromagnet on a honeycomb lattice}

\subsection{Model Hamiltonian}

We consider the $s=1/2$ quantum Heisenberg antiferromagnet on
a honeycomb lattice (see FIG.1(a)), of which the ground state is
exactly obtained. 
The Hamiltonian is given by,
$H=H_{\rm K}+H_{\rm R}$, with
\begin{eqnarray}
&&H_{\rm K}=J_1\bigl[\sum_{(ij)}\mbox{\boldmath $S$}_i\cdot 
\mbox{\boldmath $S$}_j
+\frac{1}{2}\sum_{<ij>}\mbox{\boldmath $S$}_i\cdot 
\mbox{\boldmath $S$}_j+\frac{3}{8}N  \nonumber \\
&&+\frac{2}{5}\sum_{{\scriptstyle |j-k|=|j-l|=|k-l|=\sqrt{3}}
\atop{\scriptstyle (ij)(ik)(il)}}
(\mbox{\boldmath $S$}_i\cdot 
\mbox{\boldmath $S$}_j)(\mbox{\boldmath $S$}_k\cdot 
\mbox{\boldmath $S$}_l)
\bigr]  \label{hk}\\
&&H_{\rm R}=\displaystyle{\sum_a} 
{\cal P}^{M}_a[J_2K_a+J_3V_a] \label{hr}\\
&&K_a=\frac{7}{16}
\sum_{(\mu\nu)\in \{a\}}
\mbox{\boldmath $S$}_{\mu}\cdot \mbox{\boldmath $S$}_{\nu} 
+\frac{1}{8}\sum_{<\mu\nu>\in \{a\}}
\mbox{\boldmath $S$}_{\mu}\cdot \mbox{\boldmath $S$}_{\nu} \nonumber \\
&&+\frac{1}{8}\sum_{\ll\mu\nu\gg \in \{a\}}
\mbox{\boldmath $S$}_{\mu}\cdot \mbox{\boldmath $S$}_{\nu} \nonumber \\
&&-\frac{5}{32}[\sum^a_{A}P_4^a
+\sum^a_{B}P_4^a
-\frac{3}{5}\sum^a_{C}P_4^a]
+\frac{1}{8}P_6^a+\frac{13}{64}  \\
&&V_a=-\frac{1}{16}\sum_{(\mu\nu)\in\{a\}}
\mbox{\boldmath $S$}_{\mu}\cdot \mbox{\boldmath $S$}_{\nu} \nonumber \\
&&+\frac{1}{4}\sum_{(\mu\nu)(\alpha\beta)\in \{a\}\atop \mu<\nu<\alpha<\beta}
(\mbox{\boldmath $S$}_{\mu}\cdot \mbox{\boldmath $S$}_{\nu})
(\mbox{\boldmath $S$}_{\alpha}\cdot \mbox{\boldmath $S$}_{\beta}) \nonumber \\
&&-\sum_{(\kappa\lambda)(\varepsilon\gamma)(\eta\xi)\in\{a\} \atop
(\lambda\varepsilon)(\gamma\eta)(\xi\kappa),\kappa<\varepsilon<\eta}
(\mbox{\boldmath $S$}_{\kappa}\cdot \mbox{\boldmath $S$}_{\lambda})
(\mbox{\boldmath $S$}_{\varepsilon}\cdot \mbox{\boldmath $S$}_{\gamma})
(\mbox{\boldmath $S$}_{\eta}\cdot \mbox{\boldmath $S$}_{\xi})+\frac{1}{32}  \\
&&{\cal P}_a^{M}=\frac{7}{64}+\frac{7}{48}\sum_{ij\in \{\tilde{a}\}}
\mbox{\boldmath $S$}_{i}\cdot \mbox{\boldmath $S$}_{j} \nonumber \\
&&+\frac{7}{60}\sum_{ijkl\in \{\tilde{a}\}}
(\mbox{\boldmath $S$}_{i}\cdot \mbox{\boldmath $S$}_{j})
(\mbox{\boldmath $S$}_{k}\cdot \mbox{\boldmath $S$}_{l}) \nonumber \\
&&+\frac{1}{45}\sum_{ijklmn\in \{\tilde{a}\}}
(\mbox{\boldmath $S$}_{i}\cdot \mbox{\boldmath $S$}_{j})
(\mbox{\boldmath $S$}_{k}\cdot \mbox{\boldmath $S$}_{l})
(\mbox{\boldmath $S$}_{m}\cdot \mbox{\boldmath $S$}_{n})
\end{eqnarray}
Here, $(ij)$, $<ij>$ and $\ll ij \gg$ represent, respectively, 
the nearest, next-nearest and next-next-nearest neighbor pairs, 
and $J_1\gg J_2, J_3>0$.
The summation $\sum_{\mu\nu ...\in \{a\}}$  
is taken over sites in the $a$-th hexagon.
The operators $P_4^a$ and $P_6^a$ 
represent, respectively, 
the 4-body and 6-body ring exchange interactions\cite{osh},
\begin{eqnarray}
P_4^a&=&\frac{1}{4}+\sum_{\mu<\nu}
\mbox{\boldmath $S$}_{\mu}\cdot \mbox{\boldmath $S$}_{\nu}
+4G_{ijkl}, \\
G_{ijkl}&=&(\mbox{\boldmath $S$}_{i}\cdot \mbox{\boldmath $S$}_{j})
(\mbox{\boldmath $S$}_{k}\cdot \mbox{\boldmath $S$}_{l})
+(\mbox{\boldmath $S$}_{i}\cdot \mbox{\boldmath $S$}_{l})
(\mbox{\boldmath $S$}_{j}\cdot \mbox{\boldmath $S$}_{k}) \nonumber \\
&-&(\mbox{\boldmath $S$}_{i}\cdot \mbox{\boldmath $S$}_{k})
(\mbox{\boldmath $S$}_{j}\cdot \mbox{\boldmath $S$}_{l})
\end{eqnarray}
\begin{eqnarray}
P_6^a&=&\frac{1}{16}+\frac{1}{4}\sum_{\mu<\nu}
\mbox{\boldmath $S$}_{\mu}\cdot \mbox{\boldmath $S$}_{\nu}
+\sum_{\mu<\mu<\eta<\epsilon}G_{\mu\nu\eta\epsilon}+4S_{ijklmn}, \\
S_{ijklmn}&=&\sum_{C(i,j,k,l,m,n)}(-1)^P
(\mbox{\boldmath $S$}_{\alpha}\cdot \mbox{\boldmath $S$}_{\beta})
(\mbox{\boldmath $S$}_{\gamma}\cdot \mbox{\boldmath $S$}_{\delta})
(\mbox{\boldmath $S$}_{\epsilon}\cdot \mbox{\boldmath $S$}_{\phi}), \label{sij}
\end{eqnarray}
where the sum in eq.(\ref{sij}) is taken over all combinations
of $(i,j,k,l,m,n)$ in three different pairs depicted in FIG.2.
$(-1)^P$ is $1$ (-1) when a combination is generated 
by an even (odd) number of transpositions between different pairs.
The summations $\sum_{A,B,C}^a$
are, respectively, taken
over the four-spin configurations of types A, B, and C  
of the $a$-th hexagon, as depicted in Fig.1(b).
The operator ${\cal P}_a^{M}$ projects
the six spins surrounding the $a$-th hexagon (depicted as
$i$, $j$, $k$, $l$, $m$, $n$ in Fig.1(a)) onto the subspace with
total spin $S=3$.
The summation $\sum_{ij...\in\{\tilde{a}\}}$ is taken over these six sites
in the $a$-th hexagon. 
$H_{\rm K}$ is the Hamiltonian of the $s=1/2$ Klein model 
on a honeycomb lattice,
which was studied in detail by Chayes et al.~\cite{kl,chay}
The ground state space of $H_{\rm K}$ is 
spanned by valence bond (VB) states, which are formed from
spin-singlet pairs of nearest-neighbor sites.
Thus, all dimer covering states on the honeycomb lattice are 
macroscopically degenerate ground states of $H_{\rm K}$.
An example of the dimer covering states is shown in FIG.1(c). 
The spin-spin correlation functions for these VB states exhibit
long-distance exponential decay, indicating
the absence of magnetic long-range order.
Also, the result of the single-mode approximation 
supports the existence of a spin excitation gap above the spin-singlet
ground states.
However, because each dimer state breaks the spatial symmetry 
of the system, a quantum spin liquid state is not the unique ground state of
the Klein model.

\begin{figure}[h]
\includegraphics*[width=7.5cm]{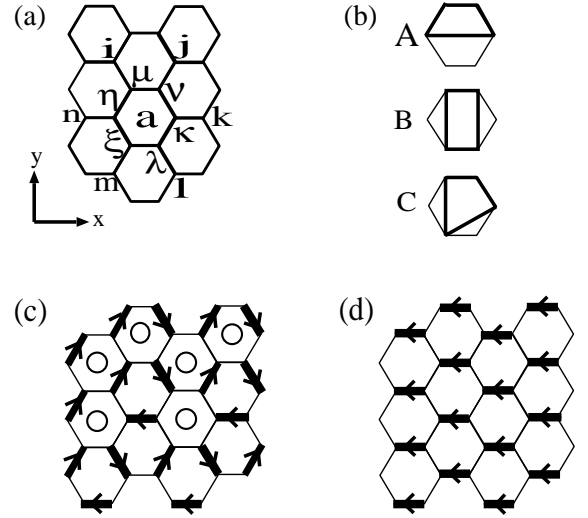}
\caption{(a) Honeycomb lattice. 
(b) Three kinds of four-spin exchange processes on a hexagon.
(c) An example of a dimer covering state. 
The arrows indicate the phase convention
of the singlet states. 
An arrow from $i$ to $j$ represents ${\cal O}_{ij}|0\rangle$.
Hexagons containing circles
are covered with unflippable dimers.
(d) Staggered VBC.
}
\end{figure}

\begin{figure}[h]
\includegraphics*[width=6cm]{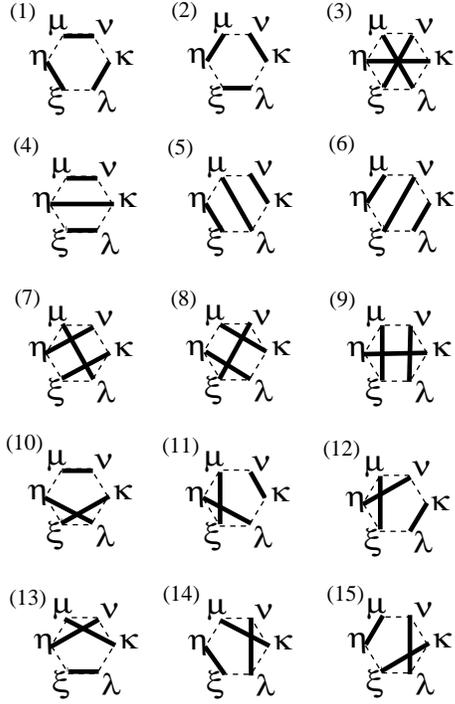}
\caption{The spin-singlet states of a hexagon.}
\end{figure}

In the following, we show that $H_{\rm R}$ introduces
a resonance among these dimer covering states
and, as a result, selects
an almost unique ground state that can be expressed as
a superposition of VB states, preserving the spatial symmetry.
To this end,  it is convenient to utilize the Schwinger boson
representation for spin operators~\cite{au}:
$S_i^{+}=u_i^{\dagger}d_i$, $S_i^{-}=d_i^{\dagger}u_i$, 
$S^z_i=(u^{\dagger}_iu_i-d^{\dagger}_id_i)/2$, and
$\hat{S}_i=(u^{\dagger}_iu_i+d^{\dagger}_id_i)/2$.
Here the average value of $\hat{S}$ is $1/2$.
We also introduce the spin-singlet operator
${\cal O}_{ij}\equiv (u_id_j-d_iu_j)/\sqrt{2}$.
Then, each dimer state 
is expressed as
\begin{eqnarray}
|D\rangle =\prod_{(ij)\in {\cal D}}{\cal O}_{ij}^{\dagger}|0\rangle. 
\end{eqnarray}
Here, ${\cal D}$ is the set of nearest-neighbor pairs corresponding to 
a particular realization of the dimer covering.
We choose the phase convention of ${\cal O}_{ij}$ as depicted in FIG.1(b).

We now show that $H_{\rm R}$ is a variant of the QDM represented
by spin operators.
The Hamiltonian of the QDM is~\cite{rk,moe2}
\begin{eqnarray}
H_{\rm QDM}=-t[| 1 \rangle \langle 2 |
+| 2 \rangle \langle 1 |] 
+V[| 1 \rangle \langle 1 |+ | 2 \rangle \langle 2 |]. 
\label{qd}
\end{eqnarray}
Here, $| 1 \rangle$ and $| 2 \rangle$ are the dimer covering
states of a single hexagon corresponding, respectively, to the states
(1) and (2) of Fig. 2.
The first term in $H_{\rm QDM}$ is the kinetic term that transfers 
$| 1 \rangle$ to $| 2 \rangle$ and $| 2 \rangle$ to $| 1 \rangle$.
The second term is the potential term.
To construct operators that act as the first and second terms
of $H_{\rm QDM}$ and are expressed in terms of spin operators, 
we consider all possible
spin singlet states of a single hexagon, depicted in Fig.2.
Although the states except (1) and (2) in Fig.3 are not the ground state of
$H_{\rm K}$, the kinetic term for the states $| 1 \rangle$ and $| 2 \rangle$ 
are expressed by the linear combination of these states.
To see this,
we introduce the density operators for the singlet dimers on a hexagon
corresponding to these fifteen states, 
\begin{eqnarray}
{\cal T}^{(1)}_a={\cal O}_{\nu\mu}^{\dagger}{\cal O}_{\nu\mu}
{\cal O}_{\lambda\kappa}^{\dagger}{\cal O}_{\lambda\kappa}
{\cal O}_{\eta\xi}^{\dagger}{\cal O}_{\eta\xi},
\end{eqnarray}
\begin{eqnarray}  
{\cal T}^{(2)}_a={\cal O}_{\nu\kappa}^{\dagger}{\cal O}_{\nu\kappa}
{\cal O}_{\lambda\xi}^{\dagger}{\cal O}_{\lambda\xi}
{\cal O}_{\eta\mu}^{\dagger}{\cal O}_{\eta\mu},
\end{eqnarray}
and so forth.
It is obvious that ${\cal T}^{(m)}| m \rangle=| m \rangle$  
with $m=1,2,...,15$.
In particular, we are concerned with the action of
${\cal T}^{(1)}_a$ and ${\cal T}^{(2)}_a$ 
on the states $| 1\rangle$ and $| 2\rangle$, i.e.
\begin{eqnarray}
&&{\cal T}^{(1)}_a| 1\rangle=| 1\rangle, \quad
{\cal T}^{(2)}_a| 2\rangle=| 2\rangle,  \nonumber \\
&&{\cal T}^{(1)}_a| 2\rangle=-\frac{1}{4}| 1\rangle, \quad
{\cal T}^{(2)}_a| 1\rangle=-\frac{1}{4}| 2\rangle. \label{tt}
\end{eqnarray}
The first pair of relations here implies 
that ${\cal T}^{(1)}_a$ and ${\cal T}^{(2)}_a$ 
act as the potential terms of $H_{\rm QDM}$.
In the following, it is convenient to use the representations
\begin{eqnarray}
{\cal T}^{(1)}_a=h^{\dagger}_{1 a}h_{1 a}, \qquad
{\cal T}^{(2)}_a=h^{\dagger}_{2 a}h_{2 a},
\end{eqnarray}
with 
\begin{eqnarray}
h_{1 a}={\cal O}_{\nu\mu}{\cal O}_{\lambda\kappa}{\cal O}_{\eta\xi}, \\
h_{2 a}={\cal O}_{\nu\kappa}{\cal O}_{\lambda\xi}{\cal O}_{\eta\mu}.
\end{eqnarray}
Next, we  introduce the operator,
\begin{eqnarray}
{\cal K}_a\equiv
-h^{\dagger}_{1 a}h_{2 a}-h^{\dagger}_{2 a}h_{1 a}, 
\end{eqnarray}
which is 
(as can be seen by using the relations 
${\cal O}_{\mu\nu}^{\dagger}{\cal O}_{\kappa\lambda}^{\dagger}
+{\cal O}_{\mu\lambda}^{\dagger}{\cal O}_{\nu\kappa}^{\dagger}
={\cal O}_{\mu\kappa}^{\dagger}{\cal O}_{\nu\lambda}^{\dagger}$, etc.)
expressed as
\begin{eqnarray}
{\cal K}_a=-\frac{1}{2}(\sum_{m=1}^2{\cal T}^{(m)}_a
-{\cal T}^{(3)}_a+\sum_{m=4}^{9}{\cal T}^{(m)}_a
-\sum_{m=10}^{15}{\cal T}^{(m)}_a).
\end{eqnarray}
Note that ${\cal K}_a$ satisfies the following relations:
\begin{eqnarray}
{\cal K}_a| 1\rangle=-| 2\rangle+\frac{1}{4}| 1\rangle, \quad 
{\cal K}_a| 2\rangle=-| 1\rangle+\frac{1}{4}| 2\rangle. \label{kk}
\end{eqnarray}
These actions of ${\cal K}_a$ 
are similar to those of the kinetic term of $H_{\rm QDM}$.
To establish the complete set of relations between the QDM and
the operators ${\cal K}_a$, ${\cal T}^{(1)}_a$ and ${\cal T}^{(2)}_a$, 
we need to verify their actions
on unflippable dimers that are not flipped by ${\cal K}_a$,
as shown in FIG.1(b).
It is easily shown that 
the actions of ${\cal K}$ and ${\cal T}^{(1)}+{\cal T}^{(2)}$ on
the unflippable dimers generate 
dimer covering states that are not in the ground-state space of 
$H_{\rm K}$. 
These non-ground states contain at least one singlet pair 
in the set of six spins surrounding the hexagon 
to which an operation is applied; i.e. 
the spins on $i$, $j$, $k$, $l$, $m$, $n$ in FIG.1(a) 
when the operators are applied to the $a$-th hexagon. 
Thus, the unwanted states are excluded by the projection
onto the maximum spin states of these six spins.
This is carried out by applying the operator ${\cal P}^{M}_a$.
Also, using the relation
$\mbox{\boldmath $S$}_i\cdot\mbox{\boldmath $S$}_j=
\frac{1}{4}-{\cal O}^{\dagger}_{ij}
{\cal O}_{ij}$,
we find ${\cal K}_a=K_a$, ${\cal T}^{(1)}_a+{\cal T}^{(2)}_a=V_a$,
thereby arriving at the Hamiltonian (\ref{hr}).

\subsection{Exact ground state}

We now show that the Hamiltonian 
$H=H_{\rm K}+H_{\rm R}$ has the same ground state properties
as the QDM in a certain parameter region.
The lowest energy sector in the Hilbert space of $H$
is spanned by VB states, provided that the ground state of $H_{\rm R}$
can be expressed as a linear combination of $|D\rangle$.
To find the ground state, it is useful to rewrite $H_{\rm R}$ into the form
\begin{eqnarray}
H_{\rm R}&=&\frac{J_2+J_3}{2}
\sum_a {\cal P}^{M}_a(h^{\dagger}_{1 a}-h^{\dagger}_{2 a})
(h_{1 a}-h_{2 a}) \nonumber \\ 
&+&\frac{J_3-J_2}{2}
\sum_a {\cal P}^{M}_a(h^{\dagger}_{1 a}+h^{\dagger}_{2 a})
(h_{1 a}+h_{2 a}).
\end{eqnarray}

For $J_3>J_2$, the energy eigenvalues of $H_{\rm R}$ are non-negative,
and only unflippable dimer states are zero energy states.
Therefore the ground state of $H$ is a dimer covering without flippable dimers.
If we simply impose periodic boundary conditions 
in both $x$ and $y$ directions, the ground state remains macroscopically 
degenerate. 
This degeneracy is removed by imposing periodic boundary conditions
with a shift by one hexagon in the $y$ direction.
Then, the unique ground state is 
the staggered VBC (see FIG.1(d)), which was previously 
identified by Moessner et al.~\cite{moe2}

The energy levels of $H_{\rm R}$ are non-negative also 
in the case of $J_2=J_3$. 
Although the staggered VBC is obviously a zero energy state here too,
in this case there also exists a non-trivial, liquid-like ground state.
It is easily seen that 
the equal-amplitude superposition of dimer states 
$|G \rangle \equiv \sum_{i} | D_i \rangle $
is the zero energy state of $H_{\rm R}$ if 
the summation $\sum_{i}$ is restricted within a sector of dimer states
related by the local operations (\ref{tt}) and (\ref{kk}).
Generally, a class of dimer states connected by these local operations
is characterized by a topological number~\cite{rk,bon}.
On a honeycomb lattice, however, in addition to a topological number, 
the total number of dimers in each direction is also conserved 
by the local flip.
Thus, $|G \rangle$ is the unique ground state of $H$
in a given sector specified by these conserved quantities.
This state, preserving both the spin-rotational and
spatial symmetries, is a spin liquid state, 
and it is in the same universality class as the short-range RVB state.
In fact, the upper limit of
the spin-spin correlation function for this state
exhibits long distance exponential decay expressed by~\cite{kl}
\begin{eqnarray}
|\langle G|S_i^zS_j^z|G\rangle |\leq (3/2)2^{-|r_i-r_j|}.
\end{eqnarray}
These results show that 
the QDM	 is realized as a spin system with multiple ring exchange 
interactions, and the case $J_2=J_3$ corresponds to
the Rokhsar-Kivelson (RK) point of the QDM~\cite{rk}.

In the case of $J_2>J_3$, unfortunately, 
we have not been able to derive an exact ground state of $H$ analytically.
However, the plaquette VBC state obtained by Moessner et al.~\cite{moe2} for
the QDM in this parameter region is indeed an exact eigenstate of $H$, and 
it is possible that this is the ground state.

We now consider the Hamiltonian $H$ 
without the projection operator ${\cal P}^{M}_a$ in $H_{\rm R}$.
Then, the model is more realistic, though its exact ground state is no longer
accessible by analytical method.
The operation of $K_a$ and $V_a$ on unflippable dimers do not vanish, but 
create states which are not in the ground state sector of $H_{\rm K}$.
In this situation, the staggered VB solid state is not the zero-energy state
even for $J_3\geq J_2$, and pushed up to a state with higher energy
of order $J_1$.
Thus, it is expected that, at the RK point $J_2=J_3$, 
the spin liquid state is the most plausible candidate 
for the true ground state provided that $J_1\gg J_2$, $J_3$.

\subsection{Dimer-dimer correlation in the spin liquid state}

In contrast to the magnetic correlation,
the dimer correlation in the spin liquid state obeys a power law,
as shown below.
In the ground state space, the dimer-dimer correlation is 
identical to
that of the classical dimer model on a honeycomb lattice, which belongs
to the universality class of the Gaussian model with
central charge $c=1~$\cite{nie}.
The correlation function for the dimer density operator 
at the bond $(ij)$,
$N_{ij}=-\mbox{\boldmath $S$}_i\cdot\mbox{\boldmath $S$}_j+1/4$, displays
the long-distance behavior
\begin{eqnarray}
\langle N_{ij}N_{lm}\rangle-\langle N_{ij}\rangle\langle N_{lm}\rangle
\sim \frac{[\cos(4\pi |r_i-r_l|/3)-1]}{|r_i-r_l|^2},
\label{dc}
\end{eqnarray}
when the two dimers on $(ij)$ and $(lm)$ are in the same direction~\cite{nie}.
This power law decay implies that the low-energy properties
are governed by gapless non-magnetic excitations. 
The low-lying excitation energy is computed using
the single-mode approximation.
We assume that the excited state takes the form
$|k \rangle=\sum_{(ij)}e^{ikr_i}\delta N_{ij}|G \rangle$,
with $\delta N_{ij}=N_{ij}-\langle N_{ij}\rangle$.
By definition, $|k \rangle$ is orthogonal to the ground state; i.e.
$\langle k| G\rangle=0$.
Therefore, the excitation energy is obtained from
\begin{eqnarray}
\varepsilon_k=\frac{\langle G|[\delta N_{-k}[H,\delta N_k]]|G\rangle}
{\langle G|\delta N_{-k}\delta N_k|G\rangle}. 
\label{sma}
\end{eqnarray}
The denominator of the right-hand side of this relation can be
computed from the correlation function $\langle N_{ij}N_{kl}\rangle$
and behaves as $\sim \ln(R/a)$, where $a$ and $R$ are
the lattice constant and the system size, respectively. 
When the excitation energy is sufficiently smaller than
the spin excitation gap of $H_{\rm K}$, 
the total number of spin-singlet dimers is conserved, and
also, $H_{\rm K}$ does not contribute to the low-lying excitations.
Thus, the numerator of the right-hand side of the above relation for 
$\varepsilon_k$ reduces to
$\sum_{il}\langle G |\delta N_{ij} H_{\rm R}\delta N_{lm}r_ir_l|G\rangle k^2$
for small $k$.
The coefficient of the $k^2$-term is expressed 
in terms of the three-body dimer correlation functions, 
which exhibits a leading logarithmic behavior of the form $\sim \ln(R/a)$.
Hence, the logarithmic divergences in the denominator and numerator
of $\varepsilon_k$ cancel out. 
We thus find that
the gapless excitation energy in the spin-singlet sector 
is given by $\varepsilon_k\sim a k^2$. 
This dispersion relation is in accordance with
Henley's conjecture based upon height representations~\cite{hen}.

\begin{figure}[h]
\includegraphics*[width=7.5cm]{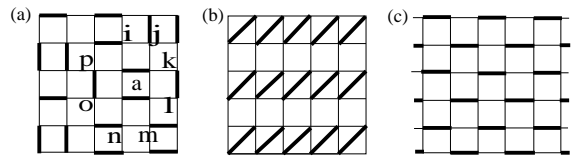}
\caption{(a) Dimer covering on a square lattice. (b) A ground state of
$H_{\rm K}$ that is not a ground state of $H_{\rm R}$. 
(c) The staggered VB crystal.
}
\end{figure}

\section{$s=1/2$ Heisenberg antiferromagnet on a square lattice}

The analysis presented in the previous sections 
is easily extended to other lattice structures.
For the model with s=1/2 on the square lattice,
the Hamiltonian is also given by $H=H_{\rm K}+H_{\rm R}$, 
but in this case we have the following:
\begin{eqnarray}
H_{\rm K}&=&J_1[\frac{1}{5}\sum_{(ij)}\mbox{\boldmath $S$}_i\cdot 
\mbox{\boldmath $S$}_j
+\frac{1}{5}\sum_{<ij>}\mbox{\boldmath $S$}_i\cdot 
\mbox{\boldmath $S$}_j
+\frac{1}{5}\sum_{\ll ij\gg}\mbox{\boldmath $S$}_i\cdot 
\mbox{\boldmath $S$}_j \nonumber \\
&+&\frac{1}{40}\sum_{{\scriptstyle i\neq j, k\neq l, i\neq k,l}
\atop{\scriptstyle j\neq k,l}} (\mbox{\boldmath $S$}_i\cdot 
\mbox{\boldmath $S$}_j)(\mbox{\boldmath $S$}_k\cdot 
\mbox{\boldmath $S$}_l)+\frac{3}{64}N], \\
H_{\rm R}&=&\displaystyle{\sum_a} 
{\cal P}^{M}_a[J_2K_a+J_3V_a], \\
K_a&=&-\frac{1}{2}\sum_{(ij)\in\{a\}}\mbox{\boldmath $S$}_i\cdot 
\mbox{\boldmath $S$}_j+\frac{1}{4}P^a_{s 4}, \\
V_a&=&\frac{1}{8}-\frac{1}{4}\sum_{(ij)\in \{a\}}\mbox{\boldmath $S$}_i\cdot 
\mbox{\boldmath $S$}_j \nonumber \\
&+&\sum_{{\scriptstyle (ij)(kl)\in \{a\}}\atop{\scriptstyle i<j<k<l}} 
(\mbox{\boldmath $S$}_i\cdot 
\mbox{\boldmath $S$}_j)(\mbox{\boldmath $S$}_k\cdot 
\mbox{\boldmath $S$}_l), \\
{\cal P}_a^{M}&=&-\frac{Q_a}{140}-\frac{Q_a^2}{280}+\frac{Q_a^3}{1260}
+\frac{Q_a^4}{2520},
\end{eqnarray}
with $Q_a=\sum_{ij\in \{\tilde{a}\}}\mbox{\boldmath $S$}_i\cdot 
\mbox{\boldmath $S$}_j$. Here the summation $\sum_{ij\in \{\tilde{a}\}}$ is 
taken over the nearest neighbor sites of the $a$-th square
(the eight sites on $i$,$j$,$k$,$l$,$m$,$n$,$o$,$p$ in FIG.3(a)).
The operator $P^a_{s 4}$ 
is the 4-body ring exchange interaction on the $a$-th square.
The ground state space of the $s=1/2$ Klein model on a square lattice 
is more complicated than that of the model on a honeycomb lattice.
In the case of a square lattice, 
in addition to dimer covering states, 
some states with spin-singlet pairs on next-nearest neighbor sites
are also in the ground state space. An example of such states is depicted
in FIG.3(b).  
However, fortunately,
such singlet states are not in the ground state space of $H_{\rm R}$,
and they can be excluded from our consideration.
It is thus found that at the RK point $J_2=J_3$, under periodic boundary 
conditions in both $x$ and $y$ directions,
the ground state of $H_{\rm K}+H_{\rm R}$ is 
the short-range RVB state $\sum_D|D\rangle$, where $|D\rangle$ is 
the dimer covering state on the square lattice.
The low energy properties of this short-range RVB state were investigated
in detail by Rokhsar and Kivelson~\cite{rk}.
For $J_3>J_2$, the staggered VBC state depicted in FIG.3(c)
is the unique ground state.

\section{$s=1$ Heisenberg antiferromagnet on a triangular lattice}

In this case, we divide the triangular lattice into three honeycomb lattices,
$h_A$, $h_B$ and $h_C$, as shown in FIG.4(a). 
Here, each site consists of two sites of two different honeycomb lattices. 
The Hamiltonian on this lattice is also given by $H=H_{\rm K}+H_{\rm R}$.
The Klein Hamiltonian $H_{\rm K}$ for this system has the same form as
that given in Eq.(\ref{hk}), but, here 
it is defined on a triangular lattice, 
and it is expressed in terms of the $s=1$ spin operators.
The ground state space of this Klein model 
is spanned by states fully packed with loops composed
of singlet dimers on the three kinds of hexagons arranged 
sequentially in order the $A$, $B$, $C$, $A$, $B$, $C$,...
We show an example of loop covering states in FIG.4.
Resonance among these loop covering states is introduced as,
\begin{eqnarray}
H_{\rm R}=H_{A}+H_{B}+H_{C}, 
\end{eqnarray}
where $H_A$, $H_B$ and $H_C$ are defined on $h_A$, $h_B$ and $h_C$, respectively, 
and take the form of Eq.(\ref{hr}) with ${\cal P}^M_a$ 
replaced by the projection onto the maximum spin state of 
the thirteen nearest neighbor spins of the $a$-th hexagon on
the parent triangular lattice:
\begin{eqnarray}
{\cal P}^M_a=\prod_{s=0}^{12}\frac{(\sum_{i\in \{\tilde{a}\}}S_i)^2-s(s+1)}
{182-s(s+1)}.
\end{eqnarray}
Here the summation  $\sum_{i\in \{\tilde{a}\}}$ is taken over
the nearest neighbor sites of the $a$-th hexagon. (See FIG.4(b).)

The ground state properties are similar to the models considered in
the previous sections.
At the RK point, the equal-amplitude superposition of loop covering states
is the unique ground state in a given sector.
The loop statistics of this state are described by a classical loop model
referred to as the ``red-green-blue model''.
The loop correlation functions of this state exhibit 
power-law long-distance behavior~\cite{ja}.
Therefore the excitation energy in the spin singlet-singlet channel 
of this system is gapless. 
This ground state represents 
a new universality class of a spin liquid described by
the quantum loop model.

\begin{figure}[h]
\includegraphics*[width=5cm]{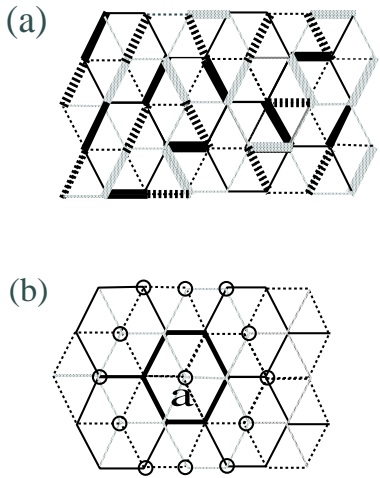}
\caption{(a) Triangular lattice divided into three honeycomb lattices, 
$h_A$~(black solid), $h_B$~(black dotted), and $h_C$~(grey).
The bold lines represent an example of loop covering.
(b) Open circles represent the nearest neighbor sites of the $a$-th hexagon.
The $a$-th hexagon is denoted by thick lines.}
\end{figure}

\section{Discussion and Summary}

In this paper, we have presented rigorous results for two-dimensional 
quantum spin systems with multiple ring exchange interactions
that possess quantum spin liquid ground states.  
It should be noted that Misguich et al. and LiMing et al. showed that 
for the $s=1/2$ Heisenberg antiferromagnet 
on a triangular lattice, sufficiently strong
ring exchange interactions destroy magnetic long-range order, 
and stabilize a spin liquid state 
~\cite{mis2,lim}.
The spin liquid state found by them has a remarkable similarity with
the ground state discussed in the present paper:
exponentially decaying spin-spin correlations, 
and a large number of spin singlet excitations in the 
spin gap (at least for some parameter regions).
Since the classical Heisenberg model with ring exchange interactions
does not possess macroscopic degeneracy,
we speculate that the underlying structure of the model may have 
a close connection with the Klein model, which is the origin of
singlet gapless excitations in our models.

These observations imply that
the realization of the spin liquid state due to 
ring exchange interactions may be generic in quantum antiferromagnets,
though the models considered in the present paper is rather complicated. 
The multiple ring exchange interaction is
a result of strong quantum fluctuations, which destabilize
conventional magnetic orders.
In real systems,
effects of such strong quantum fluctuations have been extensively studied 
so far in connection with magnetism of solid $^3$He\cite{osh}.
Also, it has been discussed that
multiple ring exchange interactions may affect significantly
the magnetic structures of NiS$_2$.
Thus, there is a possibility that the quantum spin liquid state
due to ring exchange interactions
may be realized in real systems with sufficiently strong
multiple ring exchange interactions.
 
\acknowledgements{}

The author is grateful to S. Miyashita and M. Oshikawa 
for valuable discussions.
This work was partly supported by a Grant-in-Aid from the Ministry
of Education, Science, Sports and Culture, 
Japan.

\end{document}